\documentclass[11pt,a4paper]{article}
\usepackage[utf8]{inputenc}
\usepackage{amsmath}
\usepackage{amsfonts}
\usepackage{amssymb}
\usepackage{amsbsy}
\usepackage{graphicx}
\usepackage{caption}

\usepackage{threeparttable}
\usepackage{changepage}
\usepackage{placeins}

\usepackage{tablefootnote}

\usepackage{tikz}
\usetikzlibrary{positioning,fit,arrows,shapes.geometric,chains,decorations.pathreplacing,arrows.meta}

\usepackage[super]{nth}

\usepackage{hyperref}
\hypersetup{pdfborder={0 0 1}}
\usepackage{adjustbox}

\usepackage[left=3cm,right=3cm,top=2.5cm,bottom=2.5cm]{geometry}
\usepackage{mathrsfs}
\usepackage{bm,mathtools}

\DeclareMathOperator*{\argmin}{arg\,min}

\usepackage{tikz}

\usepackage[american]{babel}
\usepackage{csquotes}

\usepackage[%
    style=numeric,
    natbib=true,
    giveninits=true,
    uniquename=init,
    alldates=year,
    backend=biber]{biblatex}
\DeclareLanguageMapping{american}{american-apa}

\newrobustcmd*{\parentexttrack}[1]{%
  \begingroup
  \blx@blxinit
  \blx@setsfcodes
  \blx@bibopenparen#1\blx@bibcloseparen
  \endgroup}

\title{\textbf{Supervised Neural Networks for Illiquid Alternative Asset Cash Flow Forecasting\thanks{We are very grateful to Satyan Malhotra, CEO at Ask2.ai, Inc as the Senior Advisor and Industry Expert. We would like to thank Annan Chen, Wen Cheng, Christopher D'Onofrio, Mateo Gomez, Ian Ho, Fatme Hourani, Nikhil Kamoji, Jiaqi Liu, Chloe Y. Moon, Curren Sameer Tipnis, Shijia Zhang, and Yongyi Zhao for their participation and help on this research. Errors are our own responsibility.}}}
\date{}
\author{Tugce Karatas\footnote{Department of IEOR, Columbia University, \textbf{tk2757@columbia.edu}} \and  Federico Klinkert\footnote{Department of IEOR, Columbia University, \textbf{fgk2106@columbia.edu}} \and  Ali Hirsa\footnote{Department of IEOR, Columbia University, \textbf{ah2347@columbia.edu}}}

\begin{document}

\maketitle
	
\begin{abstract}
Institutional investors have been increasing the allocation of the illiquid alternative assets such as private equity funds in their portfolios, yet there exists a very limited literature on cash flow forecasting of illiquid alternative assets. The net cash flow of private equity funds typically follow a J-curve pattern, however the timing and the size of the contributions and distributions depend on the investment opportunities. In this paper, we develop a benchmark model and present two novel approaches (direct vs. indirect) to predict the cash flows of private equity funds. We introduce a sliding window approach to apply on our cash flow data because different vintage year funds contain different lengths of cash flow information. We then pass the data to an \texttt{LSTM}/ \texttt{GRU} model to predict the future cash flows either directly or indirectly (based on the benchmark model). We further integrate macroeconomic indicators into our data, which allows us to consider the impact of market environment on cash flows and to apply stress testing. Our results indicate that the direct model is easier to implement compared to the benchmark model and the indirect model, but still the predicted cash flows align better with the actual cash flows. We also show that macroeconomic variables improve the performance of the direct model whereas the impact is not obvious on the indirect model. 
\end{abstract}

\providecommand{\keywords}[1]{\textbf{\textit{Keywords:}} #1}

\section{Introduction}

The portion of illiquid alternative assets, such as private equity funds, in institutional portfolios has been growing significantly as investors, mostly endowments and foundations, seek diversification relative to traditional stock and bond investments. Despite its increasing importance, so far only a few studies have attempted to understand the dynamics of cash flow forecasting in private equity investments.

\begin{figure}[!htbp]
    \centering
    \includegraphics[scale = 0.55]{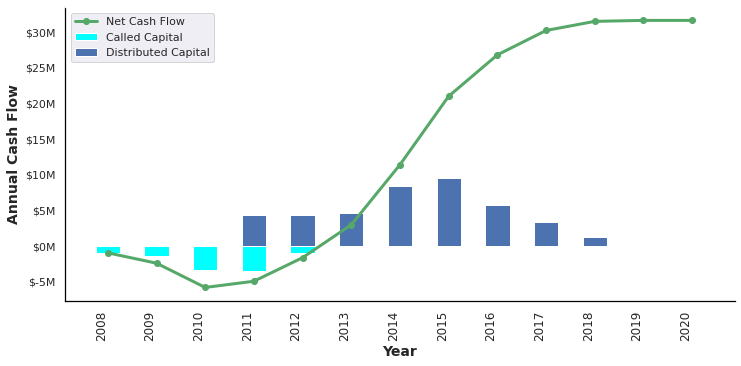}
    \caption{A Real Life Example of J-Curve }
    \label{fig:example_curve}
\end{figure}

Private equity investments are processed through limited partnerships, which are built with two parties: limited partners (LPs) and general partners (GPs). Limited partners are the investors of the private equity funds. General partners actively invest and manage the capital coming from limited partners and seek exit opportunities with sizable returns. At the onset of partnership, the investors commit a certain amount of capital that gets drawn over the life of the fund by general partners for investment purposes. In early years of a private equity fund, general partners make investments by calling capital from limited partners. Therefore, the cash flows are negative during investment period\footnote{on average, 3-5 years}. In the later years, there will be exit opportunities for the investments. When investments exit, limited partners receive distributions from general partners, and thus they obtain positive cash flows. Figure \ref{fig:example_curve} shows the evolution of cash flows of a typical private equity fund with a real life example. According to Figure \ref{fig:example_curve}, contributions are made in first five years of the fund and there are distributions to limited partners starting from the forth year. Net cash flows of private equity funds often follow a pattern similar to letter "J" as in Figure \ref{fig:example_curve}, and this pattern is known as J-curve phenomenon in private equity funds. 

Although a general pattern, J-curve, is observed in most of the private equity funds, there is an uncertainty in the timing and the size of the contributions and distributions. The schedule of contributions depends highly on general partner's investment choices, and the schedule of distributions are determined based on the exit opportunities of investments. The ambiguity in the schedule of contributions and distributions results in liquidity and cash management challenges for investors because limited partners are required to keep sufficient levels of cash whenever general partner calls for capital. Furthermore, there could be unknowable changes in the valuation of the investments made by general partners that create difficulties for predicting the future value of investments accurately. Therefore, predicting the cash flows of private equity funds is a very challenging yet a vital problem. Limited research address cash flow forecasting in private equity funds, and they all rely on a variety of assumptions on the distributions of these cash flows in their models. In this study, we aim at building an assumption-free model for cash flow forecasting by utilizing neural networks and machine learning techniques. To the best of our knowledge, there is no existing research on cash flow forecasting of illiquid products with an assumption-free model. We also utilize macroeconomic variables in order to consider the impact of market environment on the cash flow forecasting.

This paper is organized as follows: In Section \ref{LR-Yale}, we go through the current literature on cash flow forecasting in illiquid alternative assets especially in private equity funds. In Section \ref{Data-Yale}, we describe our data set and implement data analytic techniques to apply data imputation to handle missing values and data interpolation to obtain the same frequency data for all the features and labels. In Section \ref{Methodology-Yale} and Section \ref{Results-Yale}, we introduce our benchmark model and two novel approaches based on deep learning techniques, and compare the performance of the models with respect to each other and the actual cash flows. In Section \ref{Conclusion-Yale}, we summarize our results and explain our future directions.

\section{Literature Review}\label{LR-Yale}

Cash flow forecasting in illiquid alternative assets is a very challenging problem, but it is of great importance as predicting cash flows inaccurately may create cash management issues. In this section, we will go through studies which address the cash flow forecasting problem with different assumptions and methodologies.

Prior to mid-1990s, most institutional investors were pursuing strategies to increase their allocations in illiquid alternative assets. Fund valuations were estimated based on simple intuitions regarding commitment levels to the funds. However, increasing proportion of illiquid alternative assets in portfolios of institutional investors lowered the effectiveness of these intuitions on fund estimations. Some investors then started to use historical averages in cash flows by assuming future funds would follow the same pattern of capital contributions, distributions and valuations. On the other hand, historical averages may vary a lot based on the chosen time period, and it is difficult to decide on which time period to base the model. In order to overcome the problems arisen from using simple intuitions and historical averages, Takahashi and Alexander \cite{takahashi2002illiquid} introduced a discrete-time deterministic model that is simple on a theoretical basis. We will refer this model as \texttt{Yale model} for the rest of the paper, and we will use the model as the benchmark. As the cash flows are not only affected by time period but also investment environment, they initially built their model based on historical data and then made adjustments according to investment environment if necessary. In their model, there are six inputs in order to predict capital contributions, distributions, and net asset value of a private equity fund: RC (rate of contribution), CC (capital commitment), L (life of the fund), B (bow factor describing changes on the rate of distribution over time), G (annual growth rate), and Y (yield percentage). Their model consists of three deterministic equations:
\begin{align}
    C_{t} =& RC_t(CC-PIC_t),   &&\text{where } PIC_t = \sum_{0}^{t-1} C_t \label{eq1} \\ 
    D_{t} =& RD\left[NAV_{t-1}\times (1+G)\right],&&  \text{where }RD=max\left[Y,\left(t/L\right)^B\right] \label{eq2}\\
    NAV_t =& \left[NAV_{t-1}\times (1+G)\right] +C_t -D_t  \label{eq3}
\end{align}
Capital contributions are calculated by multiplying the rate of contribution and the uncalled capital that is the initial capital commitment minus paid in capital. The size of the distributions are affected by the performance of the fund significantly. Hence, distributions are modelled based on the rate of distributions and the net asset value of the fund. Equation \ref{eq3} shows that the net asset value is projected using investment performance, contributions, and distributions as predictors. The model is built upon these three simple interrelated equations, and it is adjustable with the actual data. However, the model depends on certain assumptions, and some input parameters are needed to be estimated. The values of rate of contribution, bow factor, and life of the fund assigned by investor's knowledge and growth rate is projected using the given data before implementing the model. Another drawback of the model is the dependency of contributions and distributions in the model. The calibration of the model with recent data is difficult because of this dependency.  

In order to address the issues with deterministic and interrelated model, Buchner et al. \cite{buchner2010modeling} proposed a continuous-time stochastic model with two independent components that solely relies on observable cash flow data. As the timing and the size of the contributions are unknown, they assumed that the rate of contribution ($\delta_t$) follows a mean-reverting square-root process given by the stochastic differential equation:
\begin{align}
    d\delta_t &= \kappa(\theta-\delta_t)dt+\sigma_\delta\sqrt{\delta_t}dB_{\delta,t} \label{eq4}
\end{align}
where $\theta$, $\kappa$, $\sigma_\delta$, $B_{\delta,t}$ represents the long-run mean of the rate of contribution, the rate of reversion to mean, the volatility of the rate of contribution, and standard Brownian motion, respectively. $\theta$, $\kappa$, and $\sigma_\delta$ are defined as non-negative. In order to consider the uncertainty in the distributions, they further assumed that the logarithm of instantaneous distributions( $\ln{p_t}$) follows an arithmetic Brownian motion with a time-dependent drift:
\begin{align}
    d\ln{p_t} &= \mu_t dt + \sigma_p dB_{p,t}
\end{align}
where $\mu_t$, $\sigma_p$, $B_{p,t}$ denotes the time-dependent drift, volatility, and another standard Brownian motion, respectively. They also assumed that these two Brownian motions are uncorrelated. As time-drift $\mu_t$ represents the behavior of the distributions, defining an appropriate function for $\mu_t$ is difficult. Therefore, the long-run mean of fund multiple $m$ and the speed of convergence $\alpha$ are used to obtain a good time-drift function.

Once they completed building independent stochastic processes for rate of contribution and distributions, they implemented their model empirically on a data set of 203 (102 VC and 101 buyout funds) mature European private equity funds (95 of them are liquidated) within the time horizon of January 1980 and June 2003. Since the funds have different sizes, the percentages of contributions and distributions according to the total commitment are used as input. In their model, they estimated the model parameters by using \textit{Conditional Least Squares} that is applied for estimating parameters used in continuous-time stochastic processes. They implemented the consistency tests such as Q-Q plot, Kolmogorov-Smirnov test, and coefficient of determination ($R^2$) to evaluate the performance of the model. The results show that 99.6 percent of variation in average cumulated net cash flows of the liquidated funds can be justified with the model proposed. Furthermore, this model can be used for risk management as it allows to evaluate the sensitivity of performance measures such as internal rate of return (IRR) with respect to the changes in contribution and distribution schedule. In their continuation paper, Buchner et al. \cite{buchner2014private} defined $B_{w,t}$ as Brownian motion driving aggregate stock market returns. Correlation between the rate of contributions and the market ($p_{\delta,w}$) was introduced into their model to account for the possible effect of overall stock market conditions on the rate of contributions. Similarly, they considered the effect of market on the capital distributions by introducing correlation between instantaneous distributions and the market. Overall, their models are easy to be adjusted to the recent data since contributions and distributions are modeled separately. Their models can also be used for risk analysis. However, they still rely on certain assumptions on the stochastic processes of contributions and distributions. 

As these models are approaching the problem from different angles, Furenstam and Forsell \cite{forsell2018cash} studied on the comparison of their performances with an empirical analysis. While implementing the model from Takahashi and Alexander \cite{takahashi2002illiquid}, they estimated the bow factor and the fund of the life by using the method of least squares and rate of contribution by calculating contributions and unfunded capital for the full data set of funds for each period. In order to employ the model in Buchner et al. \cite{buchner2010modeling}, they implemented conditional least squares method to estimate $\kappa$, $\theta$ and $\sigma_\delta$ parameters in contribution process, and $\alpha$ parameter in distribution process as suggested in the original paper. They calculated the long-run average multiple $m$ by taking the sample average of the fund multiples and $\sigma_p$ by considering the variance of the cumulated distributions at each discrete point in time. Once they estimated the model parameters for both studies, they employed the models on a data set that consists of 195 global buyout funds between 1999 and 2017. They evaluated their results using the coefficient of the determination ($R^2$) and backtesting the models in different historical periods. Their results indicate that the deterministic model provides better results than the stochastic model based on the data set and the estimated parameters. 

In our paper, we build an overall methodology for cash flow forecasting. We first apply cubic polynomial interpolation for imputing missing values, and then introduce sliding window approach to prepare the fund level data as input. We apply deep learning techniques such as \texttt{LSTM} and \texttt{GRU} to predict future cash flows of contributions and distributions. We also evaluate the impact of macroeconomic variables such as unemployment rate and GDP on cash flow forecasting, and build stress-testing methodologies to measure the impact of market shocks on the robustness of the model.

\section{Data}\label{Data-Yale}

The data in this study comes from Preqin, which is a popular data provider among private equity firms. Our dataset consists of the quarterly reported fund performance data of 606 North American buyout funds with a vintage year between 2000 and 2013. From Preqin, we obtain three quarterly time series for each fund: called capital (\%), distribution to paid-in (DPI \%), and residual value to paid-in (RVPI \%). Their formulas are provided in Equations \ref{called}, \ref{dist}, and \ref{residual}.

\begin{align}
    \text{Called Capital (\%)} &= \frac{\text{Total LP Contribution}}{\text{Total LP Commitment}}\times 100  \label{called}\\
    \text{Distribution to Paid-In (DPI \%)} &= \frac{\text{Total LP Distribution}}{\text{Total LP Contribution}}\times 100  \label{dist}\\
    \text{Residual Value to Paid-In (RVPI \%)} &= \frac{\text{Unrealized Value of Fund}}{\text{Total LP Contribution}}\times 100 \label{residual}
\end{align}

\noindent In Equation \ref{residual}, unrealized value of the fund means the net asset value of the fund. Contribution, distribution, and NAV of each fund are then represented as the percentage of the fund commitment level with the calculations below. Normalization of each cash flow by the commitment level allows us to predict the exact value of future cash flows by multiplying predictions with commitment level. It is important to note that CC and DC provides the ratios of cumulative contributions and cumulative distributions with respect to commitment level. In order to build a model with quarter contributions and distribution, we take the difference between consecutive CC and DC values.  

\begin{align}
    \text{Contribution to Commitment (CC)} &= \frac{\text{Capital Called (\%)}}{100}  \label{cc}\\
    \text{Distribution to Commitment (DC)} &= \frac{\text{DPI \%}\times \text{Called Capital (\%)}}{1000}  \label{dc}\\
    \text{Residual Value to Commitment (RVC)} &= \frac{\text{RVPI \%}\times \text{Called Capital (\%)}}{1000} \label{rvc}
\end{align}

\begin{figure}[!htbp]
    \hspace*{-2.0cm}
    \centering
    \includegraphics[scale = 0.52]{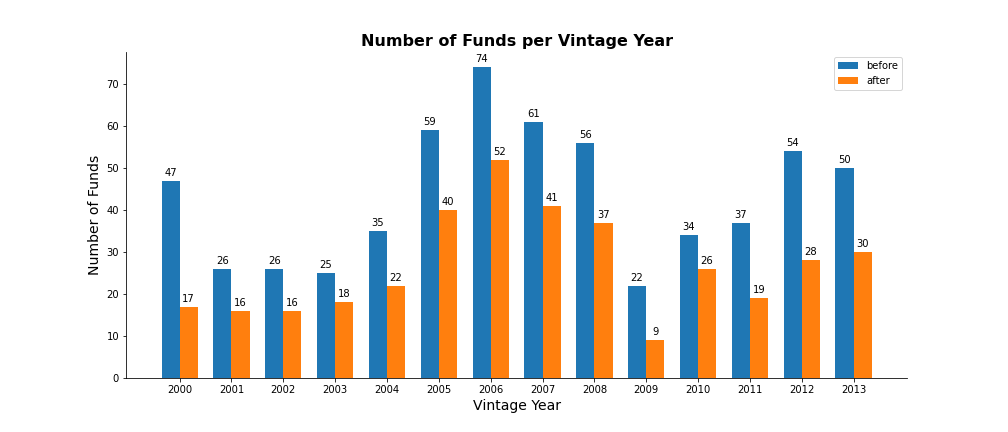}
    \caption{Number of funds by vintage year before and after removing funds with 30\% or more missing values}  
    \label{fig:vintage}
\end{figure}

We observe that cash-flow information is not provided by LPs of some funds for many quarters. For these funds, there are a lot of missing values in cash-flows, which may distort the data quality. Therefore, funds are removed from the dataset if at least one of the cash-flows has 30\% or more missing values. Figure \ref{fig:vintage} illustrates the distribution of the number of funds in our dataset corresponding to their vintage year before and after removing funds based on their missing values. The total number of the funds in our dataset reduces to 371 after this step. 

\FloatBarrier

\begin{figure}[!htbp]
   \begin{minipage}{0.5\textwidth}
   \hspace*{-1.5cm}
     \centering
     \includegraphics[scale = 0.35]{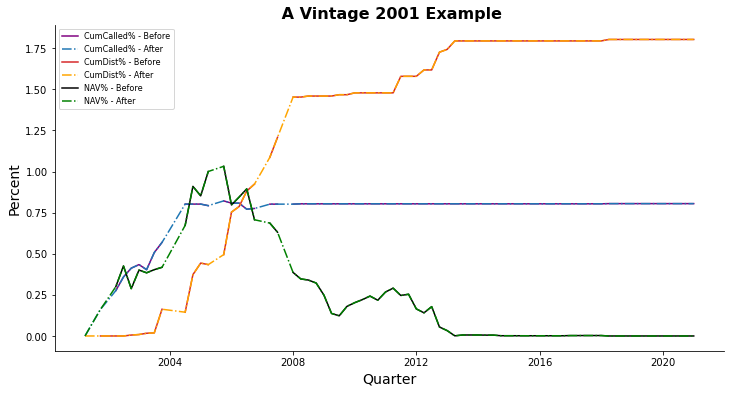}
   \end{minipage}\hfill
   \begin{minipage}{0.5\textwidth}
   \hspace*{0.5cm}
     \centering
     \includegraphics[scale = 0.35]{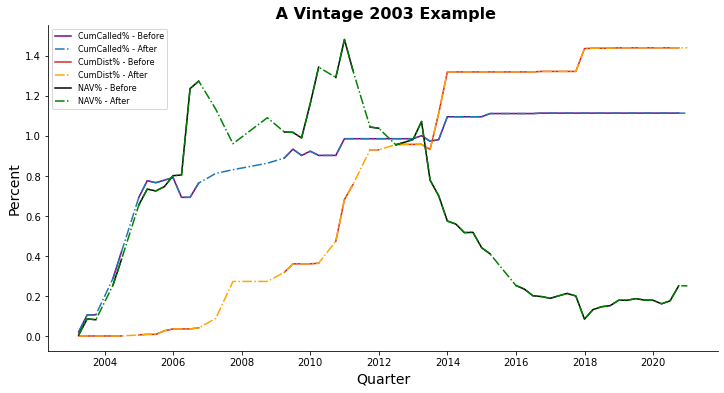}
   \end{minipage} \hfill
   \begin{minipage}{0.5\textwidth}
   \hspace*{-1.5cm}
     \centering
     \includegraphics[scale = 0.35]{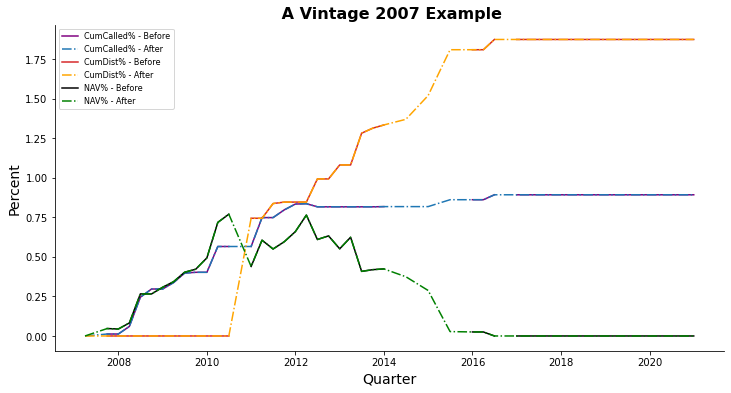}
   \end{minipage}\hfill
   \begin{minipage}{0.5\textwidth}
   \hspace*{0.5cm}
     \centering
     \includegraphics[scale = 0.35]{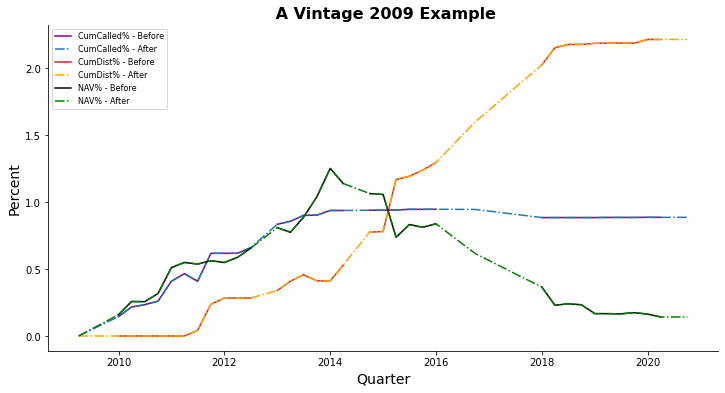}
   \end{minipage}
 \caption{Sample Plots Before/After Interpolation} \label{fig:interpolation}
\end{figure}

In order to fill the missing cash-flow values, we consider different interpolation techniques from the simplest to the most sophisticated: linear interpolation, piecewise polynomial interpolation, Brownian bridge, VG bridge, and Fourier transform techniques. Based on our empirical study, we apply piecewise cubic polynomial interpolation. Figure \ref{fig:interpolation} illustrates cash flows of four different vintage year funds before and after applying piecewise cubic polynomial interpolation. The plots show that the interpolated values for missing cash flows do not distort the expected patterns of contributions, distributions, and net asset valuations. 

Each vintage year fund in the dataset contains different length of cash flows. In order to use all data within the same model, we propose a sliding window aproach.

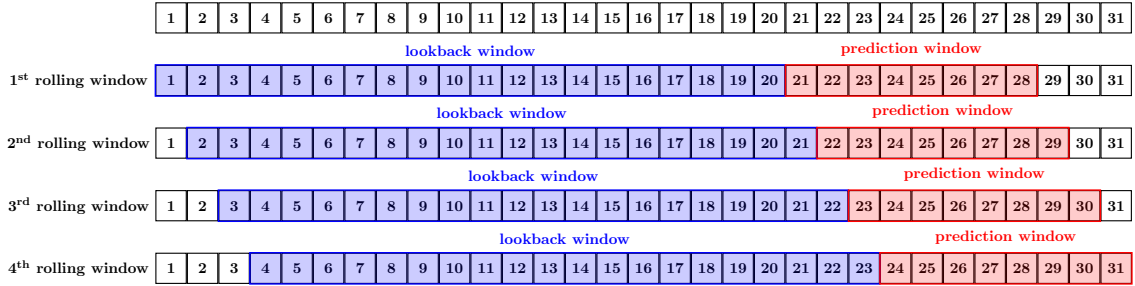
\begin{figure}[!htbp]
\centering
\begin{adjustbox}{max totalsize={1.0\textwidth}{.9\textheight},center}
\begin{tikzpicture}[auto, thick, >=triangle 45]
 \tikzset{box/.style={draw,minimum width=2em,minimum height=2em}};
\matrix [column sep={0mm}, row sep=8mm] {
    \node(n0) [color=black]{};&
    \node(n1) [box]{\textbf{1}}; & 
    \node(n2) [box]{\textbf{2}};& 
    \node(n3) [box]{\textbf{3}}; &
    \node(n4) [box]{\textbf{4}}; & 
    \node(n5) [box]{\textbf{5}};& 
    \node(n6) [box]{\textbf{6}}; &
    \node(n7) [box]{\textbf{7}}; & 
    \node(n8) [box]{\textbf{8}};& 
    \node(n9) [box]{\textbf{9}}; &
    \node(n10) [box]{\textbf{10}}; & 
    \node(n11) [box]{\textbf{11}};& 
    \node(n12) [box]{\textbf{12}}; &
    \node(n13) [box]{\textbf{13}}; & 
    \node(n14) [box]{\textbf{14}};& 
    \node(n15) [box]{\textbf{15}}; &
    \node(n16) [box]{\textbf{16}}; & 
    \node(n17) [box]{\textbf{17}};& 
    \node(n18) [box]{\textbf{18}}; &
    \node(n19) [box]{\textbf{19}}; &
    \node(n20) [box]{\textbf{20}}; & 
    \node(n21) [box]{\textbf{21}};& 
    \node(n22) [box]{\textbf{22}}; &
    \node(n23) [box]{\textbf{23}}; & 
    \node(n24) [box]{\textbf{24}};& 
    \node(n25) [box]{\textbf{25}}; &
    \node(n26) [box]{\textbf{26}}; & 
    \node(n27) [box]{\textbf{27}};& 
    \node(n28) [box]{\textbf{28}}; &
    \node(n29) [box]{\textbf{29}}; & 
    \node(n30) [box]{\textbf{30}};& 
    \node(n31) [box]{\textbf{31}}; &
    \\
    \node(n0) [color=black]{\textbf{\nth{1} rolling window}};&
    \node(n1a) [box]{\textbf{1}}; & 
    \node(n2) [box]{\textbf{2}};& 
    \node(n3) [box]{\textbf{3}}; &
    \node(n4) [box]{\textbf{4}}; & 
    \node(n5) [box]{\textbf{5}};& 
    \node(n6) [box]{\textbf{6}}; &
    \node(n7) [box]{\textbf{7}}; & 
    \node(n8) [box]{\textbf{8}};& 
    \node(n9) [box]{\textbf{9}}; &
    \node(n10) [box]{\textbf{10}}; & 
    \node(n11) [box]{\textbf{11}};& 
    \node(n12) [box]{\textbf{12}}; &
    \node(n13) [box]{\textbf{13}}; & 
    \node(n14) [box]{\textbf{14}};& 
    \node(n15) [box]{\textbf{15}}; &
    \node(n16) [box]{\textbf{16}}; & 
    \node(n17) [box]{\textbf{17}};& 
    \node(n18) [box]{\textbf{18}}; &
    \node(n19) [box]{\textbf{19}}; &
    \node(n20a) [box]{\textbf{20}}; & 
    \node(n21a) [box]{\textbf{21}};& 
    \node(n22) [box]{\textbf{22}}; &
    \node(n23) [box]{\textbf{23}}; & 
    \node(n24) [box]{\textbf{24}};& 
    \node(n25) [box]{\textbf{25}}; &
    \node(n26) [box]{\textbf{26}}; & 
    \node(n27) [box]{\textbf{27}};& 
    \node(n28a) [box]{\textbf{28}}; &
    \node(n29) [box]{\textbf{29}}; & 
    \node(n30) [box]{\textbf{30}};& 
    \node(n31) [box]{\textbf{31}}; &
    \\
    \node(n0) [color=black]{\textbf{\nth{2} rolling window}};&
    \node(n1) [box]{\textbf{1}}; & 
    \node(n2b) [box]{\textbf{2}};& 
    \node(n3) [box]{\textbf{3}}; &
    \node(n4) [box]{\textbf{4}}; & 
    \node(n5) [box]{\textbf{5}};& 
    \node(n6) [box]{\textbf{6}}; &
    \node(n7) [box]{\textbf{7}}; & 
    \node(n8) [box]{\textbf{8}};& 
    \node(n9) [box]{\textbf{9}}; &
    \node(n10) [box]{\textbf{10}}; & 
    \node(n11) [box]{\textbf{11}};& 
    \node(n12) [box]{\textbf{12}}; &
    \node(n13) [box]{\textbf{13}}; & 
    \node(n14) [box]{\textbf{14}};& 
    \node(n15) [box]{\textbf{15}}; &
    \node(n16) [box]{\textbf{16}}; & 
    \node(n17) [box]{\textbf{17}};& 
    \node(n18) [box]{\textbf{18}}; &
    \node(n19) [box]{\textbf{19}}; &
    \node(n20) [box]{\textbf{20}}; & 
    \node(n21b) [box]{\textbf{21}};& 
    \node(n22b) [box]{\textbf{22}}; &
    \node(n23) [box]{\textbf{23}}; & 
    \node(n24) [box]{\textbf{24}};& 
    \node(n25) [box]{\textbf{25}}; &
    \node(n26) [box]{\textbf{26}}; & 
    \node(n27) [box]{\textbf{27}};& 
    \node(n28) [box]{\textbf{28}}; &
    \node(n29b) [box]{\textbf{29}}; & 
    \node(n30) [box]{\textbf{30}};& 
    \node(n31) [box]{\textbf{31}}; &
    \\
    \node(n0) [color=black]{\textbf{\nth{3} rolling window}};&
    \node(n1) [box]{\textbf{1}}; & 
    \node(n2) [box]{\textbf{2}};& 
    \node(n3c) [box]{\textbf{3}}; &
    \node(n4) [box]{\textbf{4}}; & 
    \node(n5) [box]{\textbf{5}};& 
    \node(n6) [box]{\textbf{6}}; &
    \node(n7) [box]{\textbf{7}}; & 
    \node(n8) [box]{\textbf{8}};& 
    \node(n9) [box]{\textbf{9}}; &
    \node(n10) [box]{\textbf{10}}; & 
    \node(n11) [box]{\textbf{11}};& 
    \node(n12) [box]{\textbf{12}}; &
    \node(n13) [box]{\textbf{13}}; & 
    \node(n14) [box]{\textbf{14}};& 
    \node(n15) [box]{\textbf{15}}; &
    \node(n16) [box]{\textbf{16}}; & 
    \node(n17) [box]{\textbf{17}};& 
    \node(n18) [box]{\textbf{18}}; &
    \node(n19) [box]{\textbf{19}}; &
    \node(n20) [box]{\textbf{20}}; & 
    \node(n21) [box]{\textbf{21}};& 
    \node(n22c) [box]{\textbf{22}}; &
    \node(n23c) [box]{\textbf{23}}; & 
    \node(n24) [box]{\textbf{24}};& 
    \node(n25) [box]{\textbf{25}}; &
    \node(n26) [box]{\textbf{26}}; & 
    \node(n27) [box]{\textbf{27}};& 
    \node(n28) [box]{\textbf{28}}; &
    \node(n29) [box]{\textbf{29}}; & 
    \node(n30c) [box]{\textbf{30}};& 
    \node(n31) [box]{\textbf{31}}; &
    \\
    \node(n0) [color=black]{\textbf{\nth{4} rolling window}};&
    \node(n1) [box]{\textbf{1}}; & 
    \node(n2) [box]{\textbf{2}};& 
    \node(n3) [box]{\textbf{3}}; &
    \node(n4d) [box]{\textbf{4}}; & 
    \node(n5) [box]{\textbf{5}};& 
    \node(n6) [box]{\textbf{6}}; &
    \node(n7) [box]{\textbf{7}}; & 
    \node(n8) [box]{\textbf{8}};& 
    \node(n9) [box]{\textbf{9}}; &
    \node(n10) [box]{\textbf{10}}; & 
    \node(n11) [box]{\textbf{11}};& 
    \node(n12) [box]{\textbf{12}}; &
    \node(n13) [box]{\textbf{13}}; & 
    \node(n14) [box]{\textbf{14}};& 
    \node(n15) [box]{\textbf{15}}; &
    \node(n16) [box]{\textbf{16}}; & 
    \node(n17) [box]{\textbf{17}};& 
    \node(n18) [box]{\textbf{18}}; &
    \node(n19) [box]{\textbf{19}}; &
    \node(n20) [box]{\textbf{20}}; & 
    \node(n21) [box]{\textbf{21}};& 
    \node(n22) [box]{\textbf{22}}; &
    \node(n23d) [box]{\textbf{23}}; & 
    \node(n24d) [box]{\textbf{24}};& 
    \node(n25) [box]{\textbf{25}}; &
    \node(n26) [box]{\textbf{26}}; & 
    \node(n27) [box]{\textbf{27}};& 
    \node(n28) [box]{\textbf{28}}; &
    \node(n29) [box]{\textbf{29}}; & 
    \node(n30) [box]{\textbf{30}};& 
    \node(n31d) [box]{\textbf{31}}; &
    \\
};

 \tikzset{blue dotted/.style={draw=blue, fill=blue,fill opacity=0.2, line width=1pt,inner sep=0mm, rectangle}};
 
 \tikzset{red dotted/.style={draw=red,fill=red,fill opacity=0.2, line width=1pt,inner sep=0mm, rectangle}};

 \node (first dotted box) [blue dotted, fit = (n1a) (n20a)] {};
 \node (second dotted box) [red dotted,fit = (n21a) (n28a)] {};
\node at (first dotted box.north) [above, inner sep=2mm,color=blue] {\textbf{lookback window}};
\node at (second dotted box.north) [above, inner sep=2mm, color=red] {\textbf{prediction window}};

 \node (third dotted box) [blue dotted, fit = (n2b) (n21b)] {};
 \node (forth dotted box) [red dotted,fit = (n22b) (n29b)] {};
\node at (third dotted box.north) [above, inner sep=2mm,color=blue] {\textbf{lookback window}};
\node at (forth dotted box.north) [above, inner sep=2mm, color=red] {\textbf{prediction window}};

 \node (fifth dotted box) [blue dotted, fit = (n3c) (n22c)] {};
 \node (sixth dotted box) [red dotted,fit = (n23c) (n30c)] {};
\node at (fifth dotted box.north) [above, inner sep=2mm,color=blue] {\textbf{lookback window}};
\node at (sixth dotted box.north) [above, inner sep=2mm, color=red] {\textbf{prediction window}};

 \node (seventh dotted box) [blue dotted, fit = (n4d) (n23d)] {};
 \node (eight dotted box) [red dotted,fit = (n24d) (n31d)] {};
\node at (seventh dotted box.north) [above, inner sep=2mm,color=blue] {\textbf{lookback window}};
\node at (eight dotted box.north) [above, inner sep=2mm, color=red] {\textbf{prediction window}};

\end{tikzpicture}
\end{adjustbox}
\captionof{figure}{Sliding Window Approach for a 2013 Vintage Fund with $w_{in}=20$ and $w_{out}=8$.}
\label{plot:sliding_window}
\end{figure}
\FloatBarrier
We assume that we have $\mathbf{K}$ vintage years in our dataset. For each vintage year $\mathbf{k}$, there are $\mathbf{n_k}$ many funds, and the maximum length of the time series within vintage year $\mathbf{k}$ funds is $\mathbf{t_k}$. We then pad all the time series for vintage year $\mathbf{k}$ to $\mathbf{t_k}$ length. We define $\mathbf{w}$ as the rolling window of our model. Rolling window $\mathbf{w}$ is divided into two: lookback window $\mathbf{w_{in}}$ and prediction window $\mathbf{w_{out}}$. The sum of $\mathbf{w_{in}}$ and $\mathbf{w_{out}}$ is equal to rolling window $\mathbf{w}$. In our model, lookback window $\mathbf{w_{in}}$ is the number of data points to be used in predicting the cash flows in prediction window $\mathbf{w_{out}}$. As the length of the time series for each fund in vintage year $\mathbf{k}$ is $\mathbf{t_k}$, the number of rolling windows for a vintage year $\mathbf{k}$ fund then becomes $\mathbf{t_k-w+1}$. Figure \ref{plot:sliding_window} illustrates sliding window approach for a 2013 vintage year fund with $\mathbf{w_{in}=20}$ and $\mathbf{w_{out}=8}$. Accordingly, we obtain four rolling windows for each 2013 vintage fund. Our overall dataset consists of $\sum_{k=1}^K (t_k-w+1)n_k$ observations that each observation includes a lookback window $\mathbf{w_{in}}$ and prediction window $\mathbf{w_{out}}$.

\section{Methodology}\label{Methodology-Yale}
Our dataset consists of North American buyout private equity funds for different vintage years. Therefore, the length of the time series differs based on the vintage year of the fund. In this section, we first introduce our benchmark model. We then propose two different models, where we use the sliding window approach to consider different length of time series, and then predict a certain number of quarters ahead by employing \textit{Recurrent Neural Networks} (RNN) models. Recurrent neural networks hold an advantage over many machine learning algorithms and feedforward neural networks because they learn the temporal relationships within the data. Both models read the multi-dimensional cash-flows one at a time from beginning to the end of the lookback window, and then they project the cash-flows in the future either directly or using a parametric approach. In our study, we specifically consider \textit{Long Short-Term Memory} (LSTM) and \textit{Gated Recurrent Unit} (GRU) models since they can remember longer-term dependencies within the time series of cash flows. Sliding window approach allows us to use different vintage year funds within a single model, but it also create a bias towards old vintage year funds since older vintage funds possess more rolling windows. To overcome the bias problem, we weight the rolling windows of each fund with the reciprocal of the total number of rolling windows that fund owns.

\subsection{Benchmark Model} \label{benchmark}
As mentioned in Section \ref{LR-Yale}, we implement Yale model as a benchmark in order to compare the model performance. However, we need to make adjustments to Yale model because Yale model is constructed using annual cash flow data, and our dataset consists of quarterly ratios. Equations \ref{eq1_1}, \ref{eq2_1}, and \ref{eq3_1} illustrate the quarterly adjustment of Yale model on contributions, distributions, and NAV that are represented as the percentage of the fund commitment level. CC, DC, and RVC are already defined in Equations \ref{called}, \ref{dist}, and \ref{residual}. qCC and qDC represent the quarterly contribution and distribution ratios. For quarterly ratios, we divide our predictions by four, and we take the power of 0.25 for the RVC. 

\begin{align}
    \widehat{qCC_{t}} =& \frac{RC_t(1-CC_{t-1})}{4} \label{eq1_1} \\ 
    \widehat{qDC_{t}} =& \frac{max\left[Y,\left(t/L\right)^B\right]\left[RVC_{t-1}\times (1+G)^{0.25}\right]}{4} \label{eq2_1}\\
    \widehat{RVC_t} =& \left[RVC_{t-1}\times (1+G)^{0.25}\right] +\widehat{qCC_{t}} -\widehat{qDC_{t}}  \label{eq3_1}
\end{align}

In order to use the benchmark model, we also need to estimate the parameters of life of the fund ($L$), bow factor($B$), annual growth rate ($G$), and rate of contribution ($RC$). Considering the industry average for the life of the fund, we assume it is equal to 16 years. As rate of contribution is only used in quarterly contributions, we estimate RC by minimizing sum of squared errors over the look ahead window $\mathbf{w_{out}}$. We also constrain the value of RC between 0 and 1. Equation \ref{eq:op1} illustrates the optimization problem. 

\begin{gather}
  \widehat{RC} \coloneqq
  \argmin_{0\leq RC \leq 1}
  \Bigl\{
    \sum_{t=1}^{\mathbf{w_{out}}} \left(qCC_t-\widehat{qCC}_t\right)^2
  \Bigr\}   \label{eq:op1}
\end{gather}
Equation \ref{eq:opt2} shows the optimization routine for estimating annual growth rate and bow factor. We minimize the sum of squared errors of distributions and net asset values to find growth rate and bow factor simultaneously. 
\begin{gather}
  \widehat{G}, \widehat{B} \coloneqq
  \argmin_{G,B}
  \Bigl\{
    \sum_{t=1}^{\mathbf{w_{out}}} \left(qDC_t-\widehat{qDC}_t\right)^2 + \sum_{t=1}^{\mathbf{w_{out}}} \left(RVC_t-\widehat{RVC}_t\right)^2 : 0\leq B\leq 5; 0\leq G \leq 1
  \Bigr\}   \label{eq:opt2}
\end{gather}
We implement these two optimization routines to estimate RC, G, and B parameters for each fund and each sliding window.

\subsection{Indirect Cash Flow Forecasting Model} \label{indirect}
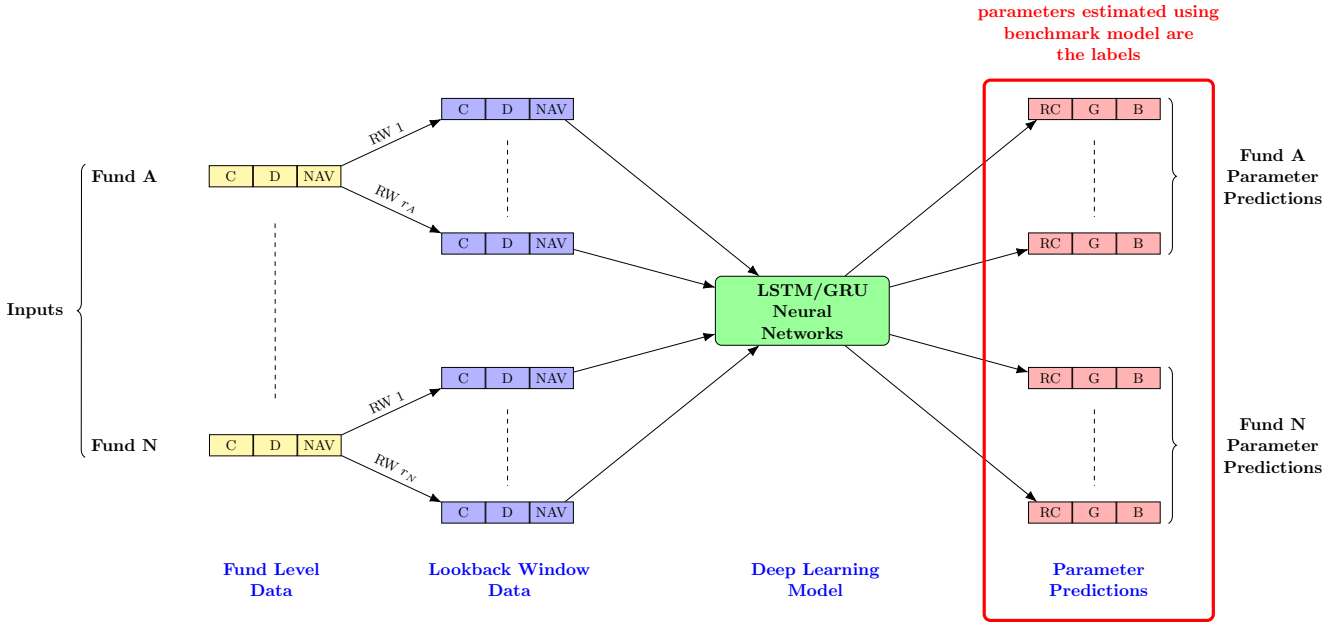
\begin{figure}[!htbp]
\centering
\begin{adjustbox}{max totalsize={1.2\textwidth}{1.2\textheight},center}
\begin{tikzpicture}[
start chain=4,node distance=5mm
]
\node[on chain=4,draw,inner sep=2pt,minimum height=4em,minimum width=10em,rounded corners,fill=green!40] at (15,0cm) 
  (x4) {\parbox{2cm}{\centering \textbf{LSTM/GRU} \\ \textbf{Neural Networks}}};
 \begin{scope}[start chain=2]
\node[on chain=2] at (0,3cm) 
  (x2) {\textbf{Fund A}};
\node[on chain=2,minimum width=2.5em,draw,right=1cm of x2,font=\footnotesize,fill=yellow!40] 
  (w2) {C};
\node[on chain=2,minimum width=2.5em,draw,right=0cm of w2,font=\footnotesize,fill=yellow!40] 
  (w21) {D};
 \node[on chain=2,minimum width=2.5em,draw,right=0cm of w21,font=\footnotesize,fill=yellow!40]  
  (w22) {NAV};
\end{scope}
 \begin{scope}[start chain=6]
\node[on chain=6] at (0,-3cm) 
  (x6) {\textbf{Fund N}};
\node[on chain=6,minimum width=2.5em,draw,right=1cm of x6,font=\footnotesize,fill=yellow!40]   
  (w6) {C};
\node[on chain=6,minimum width=2.5em,draw,right=0cm of w6,font=\footnotesize,fill=yellow!40]   
  (w61) {D};
 \node[on chain=6,minimum width=2.5em,draw,right=0cm of w61,font=\footnotesize,fill=yellow!40]   
  (w62) {NAV};
\end{scope}
 \begin{scope}[start chain=1]
\node[on chain=1,minimum width=2.5em,draw,font=\footnotesize,fill=blue!30]  at (7.5,4.5cm) 
  (w1) {C};
\node[on chain=1,minimum width=2.5em,draw,right=0cm of w1,font=\footnotesize,fill=blue!30]  
  (w11) {D};
 \node[on chain=1,minimum width=2.5em,draw,right=0cm of w11,font=\footnotesize,fill=blue!30] 
  (w12) {NAV};
\node[on chain=1,minimum width=2.5em,draw,font=\footnotesize,fill=red!30]  at (19.5,4.5cm) 
  (w1a) {RC};
\node[on chain=1,minimum width=2.5em,draw,right=0cm of w1a,font=\footnotesize,fill=red!30]  
  (w11a) {G};
 \node[on chain=1,minimum width=2.5em,draw,right=0cm of w11a,font=\footnotesize,fill=red!30] 
  (w12a) {B};
\end{scope}

 \begin{scope}[start chain=3]
\node[on chain=3,minimum width=2.5em,draw,font=\footnotesize,fill=blue!30]   at (7.5,1.5cm) 
  (w3) {C};
\node[on chain=3,minimum width=2.5em,draw,right=0cm of w3,font=\footnotesize,fill=blue!30]  
  (w31) {D};
 \node[on chain=3,minimum width=2.5em,draw,right=0cm of w31,font=\footnotesize,fill=blue!30]  
  (w32) {NAV};
\node[on chain=3,minimum width=2.5em,draw,font=\footnotesize,fill=red!30] at (19.5,1.5cm)  
  (w3a) {RC};
\node[on chain=3,minimum width=2.5em,draw,right=0cm of w3a,font=\footnotesize,fill=red!30]  
  (w31a) {G};
 \node[on chain=3,minimum width=2.5em,draw,right=0cm of w31a,font=\footnotesize,fill=red!30]  
  (w32a) {B};
\end{scope}

 \begin{scope}[start chain=5]
\node[on chain=5,minimum width=2.5em,draw,font=\footnotesize,fill=blue!30] at (7.5,-1.5cm) 
  (w5) {C};
\node[on chain=5,minimum width=2.5em,draw,right=0cm of w5,font=\footnotesize,fill=blue!30]  
  (w51) {D};
 \node[on chain=5,minimum width=2.5em,draw,right=0cm of w51,font=\footnotesize,fill=blue!30]  
  (w52) {NAV};
\node[on chain=5,minimum width=2.5em,draw,font=\footnotesize,fill=red!30]  at (18,-1.5cm)  at (19.5,-1.5cm) 
  (w5a) {RC};
\node[on chain=5,minimum width=2.5em,draw,right=0cm of w5a,font=\footnotesize,fill=red!30] 
  (w51a) {G};
 \node[on chain=5,minimum width=2.5em,draw,right=0cm of w51a,font=\footnotesize,fill=red!30] 
  (w52a) {B};
\end{scope}

 \begin{scope}[start chain=7]
\node[on chain=7,minimum width=2.5em,draw,font=\footnotesize,fill=blue!30] at (7.5,-4.5cm) 
  (w7) {C};
\node[on chain=7,minimum width=2.5em,draw,right=0cm of w7,font=\footnotesize,fill=blue!30]  
  (w71) {D};
 \node[on chain=7,minimum width=2.5em,draw,right=0cm of w71,font=\footnotesize,fill=blue!30] 
  (w72) {NAV};
\node[on chain=7,minimum width=2.5em,draw,font=\footnotesize,fill=red!30] at (19.5,-4.5cm)  
  (w7a) {RC};
\node[on chain=7,minimum width=2.5em,draw,right=0cm of w7a,font=\footnotesize,fill=red!30]  
  (w71a) {G};
 \node[on chain=7,minimum width=2.5em,draw,right=0cm of w71a,font=\footnotesize,fill=red!30] 
  (w72a) {B};
\end{scope}

 \begin{scope}[start chain=8]
\node[on chain=8,color=blue] at (3.25,-6cm) 
(t8) {\parbox{3cm}{\centering \textbf{Fund Level} \\ \textbf{Data}}};
\node[on chain=8,right=1.5cm of t8,color=blue]  
  (t81) {\parbox{4cm}{\centering \textbf{Lookback Window} \\ \textbf{Data}}};
\node[on chain=8,right=2.5cm of t81,color=blue]  
  (t82) {\parbox{4cm}{\centering \textbf{Deep Learning} \\ \textbf{Model}}};
\node[on chain=8,right=2cm of t82,color=blue]  
  (t83) {\parbox{4cm}{\centering \textbf{Parameter} \\ \textbf{Predictions}}};
\end{scope}

\draw[dashed,dash phase=3pt,inner sep=2pt,shorten >=20pt,shorten <=20pt] (w21) -- (w61);
\draw[dashed,dash phase=3pt,inner sep=2pt,shorten >=10pt,shorten <=10pt] (w11) -- (w31);
\draw[dashed,dash phase=3pt,inner sep=2pt,shorten >=10pt,shorten <=10pt] (w51) -- (w71);
\draw[dashed,dash phase=3pt,inner sep=2pt,shorten >=10pt,shorten <=10pt] (w11a) -- (w31a);
\draw[dashed,dash phase=3pt,inner sep=2pt,shorten >=10pt,shorten <=10pt] (w51a) -- (w71a);
\draw[-{Latex[length=2.5mm]}] (w22) -- (w1) node[font=\footnotesize,sloped,above,pos=0.5] {RW 1};
\draw[-{Latex[length=2.5mm]}] (w22) -- (w3)node[font=\footnotesize,sloped,above,pos=0.5] {RW $r_A$};
\draw[-{Latex[length=2.5mm]}] (w62) -- (w5)node[font=\footnotesize,sloped,above,pos=0.5] {RW 1};
\draw[-{Latex[length=2.5mm]}] (w62) -- (w7)node[font=\footnotesize,sloped,above,pos=0.5] {RW $r_N$};
\draw[-{Latex[length=2.5mm]}] (w12) -- (x4);
\draw[-{Latex[length=2.5mm]}] (w32) -- (x4);
\draw[-{Latex[length=2.5mm]}] (w52) -- (x4);
\draw[-{Latex[length=2.5mm]}] (w72) -- (x4);
\draw[-{Latex[length=2.5mm]}] (x4) -- (w1a);
\draw[-{Latex[length=2.5mm]}] (x4) -- (w3a);
\draw[-{Latex[length=2.5mm]}] (x4) -- (w5a);
\draw[-{Latex[length=2.5mm]}] (x4) -- (w7a);
\draw[decorate,decoration={brace,mirror,amplitude=5pt}] (x2.north west) -- node[left=10pt] {\textbf{Inputs}} (x6.south west);
\draw[decorate,decoration={brace,amplitude=5pt,raise=5pt}] (w12a.north east) -- node[right=10pt] {\parbox{4cm}{\centering \textbf{Fund A} \\ \textbf{Parameter} \\ \textbf{Predictions}}} (w32a.south east);
\draw[decorate,decoration={brace,amplitude=5pt,raise=5pt}] (w52a.north east) -- node[right=10pt] {\parbox{4cm}{\centering \textbf{Fund N} \\ \textbf{Parameter} \\ \textbf{Predictions}}} (w72a.south east);

 \tikzset{blue dotted/.style={draw=red, line width=2pt,inner sep=4mm, rectangle, rounded corners}};
 
 \node (first dotted box) [blue dotted, fit = (w1a)  (w12a) (t83)] {};
\node at (first dotted box.north) [above, inner sep=4mm,color=red] {\parbox{6cm}{\centering \textbf{parameters estimated using} \\ \textbf{benchmark model are} \\ \textbf{the labels}}};

\end{tikzpicture}
\end{adjustbox}
\captionof{figure}{Private Equity Cash Flow Forecasting - Indirect Model}
\label{plot:pe_framework_1}
\end{figure}

In our preliminary model, we predict future cash flows based on Yale model. Using sliding window approach, we first convert fund level data into rolling window data because each vintage year fund has different length of cash flows data. For each rolling window, we use $\mathbf{w_{out}}$ ahead cash flows to estimate RC, G, and B parameters by applying optimization routines \ref{eq:op1} and \ref{eq:opt2} in Section \ref{benchmark}. The estimated RC, G, and B parameters are then used as output data for neural networks. Figure \ref{plot:pe_framework_1} illustrates the private equity cash flow forecasting model using indirect approach. Figure \ref{plot:pe_framework_1} shows that fund level data is first converted into rolling window data. For each rolling window and each fund, we have corresponding RC, G, and B parameters as output. Therefore, we feed rolling window data into an \texttt{LSTM/GRU} model, and predict RC, G, and B parameters to calculate $\mathbf{w_{out}}$ ahead predictions using Equations \ref{eq1_1}, \ref{eq2_1}, and \ref{eq3_1}.

\FloatBarrier
\subsection{Direct Cash Flow Forecasting Model} \label{direct}

In this model, future cash flows are predicted directly without relying on any assumptions. Once rolling windows are prepared, multi-dimensional cash-flow observations for each fund and each lookback window are fed into an \texttt{LSTM/GRU model}. The model then outputs $\mathbf{w_{out}}$ ahead predictions for cash flows of the corresponding fund and rolling window. 
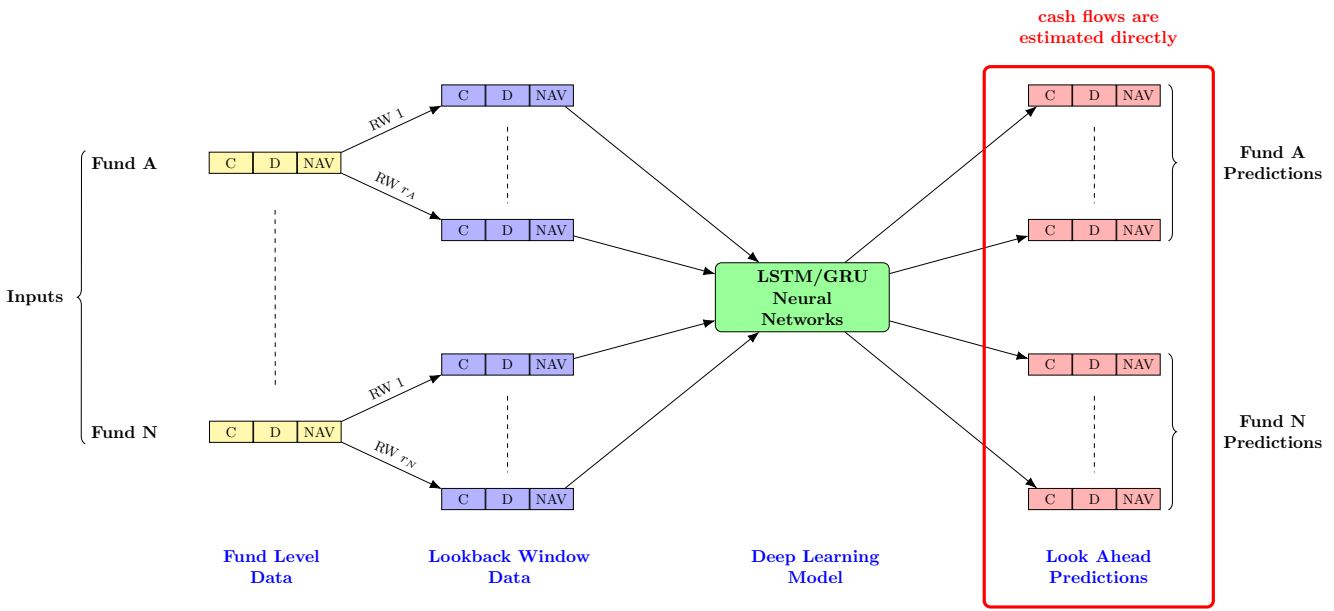
\begin{figure}[!htbp]
\centering
\begin{adjustbox}{max totalsize={1.2\textwidth}{1.2\textheight},center}
\begin{tikzpicture}[
start chain=4,node distance=5mm
]
\node[on chain=4,draw,inner sep=2pt,minimum height=4em,minimum width=10em,rounded corners,fill=green!40] at (15,0cm) 
  (x4) {\parbox{2cm}{\centering \textbf{LSTM/GRU} \\ \textbf{Neural Networks}}};
 \begin{scope}[start chain=2]
\node[on chain=2] at (0,3cm) 
  (x2) {\textbf{Fund A}};
\node[on chain=2,minimum width=2.5em,draw,right=1cm of x2,font=\footnotesize,fill=yellow!40] 
  (w2) {C};
\node[on chain=2,minimum width=2.5em,draw,right=0cm of w2,font=\footnotesize,fill=yellow!40] 
  (w21) {D};
 \node[on chain=2,minimum width=2.5em,draw,right=0cm of w21,font=\footnotesize,fill=yellow!40]  
  (w22) {NAV};
\end{scope}
 \begin{scope}[start chain=6]
\node[on chain=6] at (0,-3cm) 
  (x6) {\textbf{Fund N}};
\node[on chain=6,minimum width=2.5em,draw,right=1cm of x6,font=\footnotesize,fill=yellow!40]   
  (w6) {C};
\node[on chain=6,minimum width=2.5em,draw,right=0cm of w6,font=\footnotesize,fill=yellow!40]   
  (w61) {D};
 \node[on chain=6,minimum width=2.5em,draw,right=0cm of w61,font=\footnotesize,fill=yellow!40]   
  (w62) {NAV};
\end{scope}
 \begin{scope}[start chain=1]
\node[on chain=1,minimum width=2.5em,draw,font=\footnotesize,fill=blue!30]  at (7.5,4.5cm) 
  (w1) {C};
\node[on chain=1,minimum width=2.5em,draw,right=0cm of w1,font=\footnotesize,fill=blue!30]  
  (w11) {D};
 \node[on chain=1,minimum width=2.5em,draw,right=0cm of w11,font=\footnotesize,fill=blue!30] 
  (w12) {NAV};
\node[on chain=1,minimum width=2.5em,draw,font=\footnotesize,fill=red!30]  at (19.5,4.5cm) 
  (w1a) {C};
\node[on chain=1,minimum width=2.5em,draw,right=0cm of w1a,font=\footnotesize,fill=red!30]  
  (w11a) {D};
 \node[on chain=1,minimum width=2.5em,draw,right=0cm of w11a,font=\footnotesize,fill=red!30] 
  (w12a) {NAV};
\end{scope}

 \begin{scope}[start chain=3]
\node[on chain=3,minimum width=2.5em,draw,font=\footnotesize,fill=blue!30]   at (7.5,1.5cm) 
  (w3) {C};
\node[on chain=3,minimum width=2.5em,draw,right=0cm of w3,font=\footnotesize,fill=blue!30]  
  (w31) {D};
 \node[on chain=3,minimum width=2.5em,draw,right=0cm of w31,font=\footnotesize,fill=blue!30]  
  (w32) {NAV};
\node[on chain=3,minimum width=2.5em,draw,font=\footnotesize,fill=red!30] at (19.5,1.5cm)  
  (w3a) {C};
\node[on chain=3,minimum width=2.5em,draw,right=0cm of w3a,font=\footnotesize,fill=red!30]  
  (w31a) {D};
 \node[on chain=3,minimum width=2.5em,draw,right=0cm of w31a,font=\footnotesize,fill=red!30]  
  (w32a) {NAV};
\end{scope}

 \begin{scope}[start chain=5]
\node[on chain=5,minimum width=2.5em,draw,font=\footnotesize,fill=blue!30] at (7.5,-1.5cm) 
  (w5) {C};
\node[on chain=5,minimum width=2.5em,draw,right=0cm of w5,font=\footnotesize,fill=blue!30]  
  (w51) {D};
 \node[on chain=5,minimum width=2.5em,draw,right=0cm of w51,font=\footnotesize,fill=blue!30]  
  (w52) {NAV};
\node[on chain=5,minimum width=2.5em,draw,font=\footnotesize,fill=red!30]  at (18,-1.5cm)  at (19.5,-1.5cm) 
  (w5a) {C};
\node[on chain=5,minimum width=2.5em,draw,right=0cm of w5a,font=\footnotesize,fill=red!30] 
  (w51a) {D};
 \node[on chain=5,minimum width=2.5em,draw,right=0cm of w51a,font=\footnotesize,fill=red!30] 
  (w52a) {NAV};
\end{scope}

 \begin{scope}[start chain=7]
\node[on chain=7,minimum width=2.5em,draw,font=\footnotesize,fill=blue!30] at (7.5,-4.5cm) 
  (w7) {C};
\node[on chain=7,minimum width=2.5em,draw,right=0cm of w7,font=\footnotesize,fill=blue!30]  
  (w71) {D};
 \node[on chain=7,minimum width=2.5em,draw,right=0cm of w71,font=\footnotesize,fill=blue!30] 
  (w72) {NAV};
\node[on chain=7,minimum width=2.5em,draw,font=\footnotesize,fill=red!30] at (19.5,-4.5cm)  
  (w7a) {C};
\node[on chain=7,minimum width=2.5em,draw,right=0cm of w7a,font=\footnotesize,fill=red!30]  
  (w71a) {D};
 \node[on chain=7,minimum width=2.5em,draw,right=0cm of w71a,font=\footnotesize,fill=red!30] 
  (w72a) {NAV};
\end{scope}

 \begin{scope}[start chain=8]
\node[on chain=8,color=blue] at (3.25,-6cm) 
(t8) {\parbox{3cm}{\centering \textbf{Fund Level} \\ \textbf{Data}}};
\node[on chain=8,right=1.5cm of t8,color=blue]  
  (t81) {\parbox{4cm}{\centering \textbf{Lookback Window} \\ \textbf{Data}}};
\node[on chain=8,right=2.5cm of t81,color=blue]  
  (t82) {\parbox{4cm}{\centering \textbf{Deep Learning} \\ \textbf{Model}}};
\node[on chain=8,right=2cm of t82,color=blue]  
  (t83) {\parbox{4cm}{\centering \textbf{Look Ahead} \\ \textbf{Predictions}}};
\end{scope}

\draw[dashed,dash phase=3pt,inner sep=2pt,shorten >=20pt,shorten <=20pt] (w21) -- (w61);
\draw[dashed,dash phase=3pt,inner sep=2pt,shorten >=10pt,shorten <=10pt] (w11) -- (w31);
\draw[dashed,dash phase=3pt,inner sep=2pt,shorten >=10pt,shorten <=10pt] (w51) -- (w71);
\draw[dashed,dash phase=3pt,inner sep=2pt,shorten >=10pt,shorten <=10pt] (w11a) -- (w31a);
\draw[dashed,dash phase=3pt,inner sep=2pt,shorten >=10pt,shorten <=10pt] (w51a) -- (w71a);
\draw[-{Latex[length=2.5mm]}] (w22) -- (w1) node[font=\footnotesize,sloped,above,pos=0.5] {RW 1};
\draw[-{Latex[length=2.5mm]}] (w22) -- (w3)node[font=\footnotesize,sloped,above,pos=0.5] {RW $r_A$};
\draw[-{Latex[length=2.5mm]}] (w62) -- (w5)node[font=\footnotesize,sloped,above,pos=0.5] {RW 1};
\draw[-{Latex[length=2.5mm]}] (w62) -- (w7)node[font=\footnotesize,sloped,above,pos=0.5] {RW $r_N$};
\draw[-{Latex[length=2.5mm]}] (w12) -- (x4);
\draw[-{Latex[length=2.5mm]}] (w32) -- (x4);
\draw[-{Latex[length=2.5mm]}] (w52) -- (x4);
\draw[-{Latex[length=2.5mm]}] (w72) -- (x4);
\draw[-{Latex[length=2.5mm]}] (x4) -- (w1a);
\draw[-{Latex[length=2.5mm]}] (x4) -- (w3a);
\draw[-{Latex[length=2.5mm]}] (x4) -- (w5a);
\draw[-{Latex[length=2.5mm]}] (x4) -- (w7a);
\draw[decorate,decoration={brace,mirror,amplitude=5pt}] (x2.north west) -- node[left=10pt] {\textbf{Inputs}} (x6.south west);
\draw[decorate,decoration={brace,amplitude=5pt,raise=5pt}] (w12a.north east) -- node[right=10pt] {\parbox{4cm}{\centering \textbf{Fund A} \\ \textbf{Predictions}}} (w32a.south east);
\draw[decorate,decoration={brace,amplitude=5pt,raise=5pt}] (w52a.north east) -- node[right=10pt] {\parbox{4cm}{\centering \textbf{Fund N} \\ \textbf{Predictions}}} (w72a.south east);

 \tikzset{blue dotted/.style={draw=red, line width=2pt,inner sep=4mm, rectangle, rounded corners}};
 
 \node (first dotted box) [blue dotted, fit = (w1a)  (w12a) (t83)] {};
\node at (first dotted box.north) [above, inner sep=4mm,color=red] {\parbox{4cm}{\centering \textbf{cash flows are} \\ \textbf{estimated directly} }};

\end{tikzpicture}
\end{adjustbox}
\captionof{figure}{Private Equity Cash Flow Forecasting - Direct Model}
\label{plot:pe_framework}
\end{figure}
\FloatBarrier

Figure \ref{plot:pe_framework} shows our overall framework for private equity cash flow forecasting. Accordingly, fund level data is first converted into lookback window data for each fund and each rolling window, and then the lookback window data is passed to a chosen \texttt{RNN} model to predict future cash-flows.

\section{Results}\label{Results-Yale}
In this section, we experiment with different input data, output data and modeling architectures. We first divide the set of experiments into two: indirect cash flow forecasting model and direct cash flow forecasting model. For direct cash flow forecasting model, we also build single vintage year models and multi-vintage year models. For each model, we determine the architecture and then tune hyperparameters using Bayesian optimization techniques. Lookback window and prediction window for each model are determined as 20 quarters and 8 quarters, respectively. Therefore, we estimate 2 years ahead cash-flow predictions by utilizing 5 years cash-flow information. We select 80\% of the funds from each vintage year randomly as our training dataset, and use remaining funds for validating the models. Since we have more rolling windows for older vintage year funds, we weight each fund by the reciprocal of the number of rolling windows, which will put more importance on the recent funds. 

\subsection{Indirect Cash Flow Forecasting Results}
\begin{figure}[!htbp]
\captionsetup{font=scriptsize,labelfont=scriptsize}
   \begin{minipage}{0.5\textwidth}
   \hspace*{-1.5cm}
     \centering
     \includegraphics[scale = 0.35]{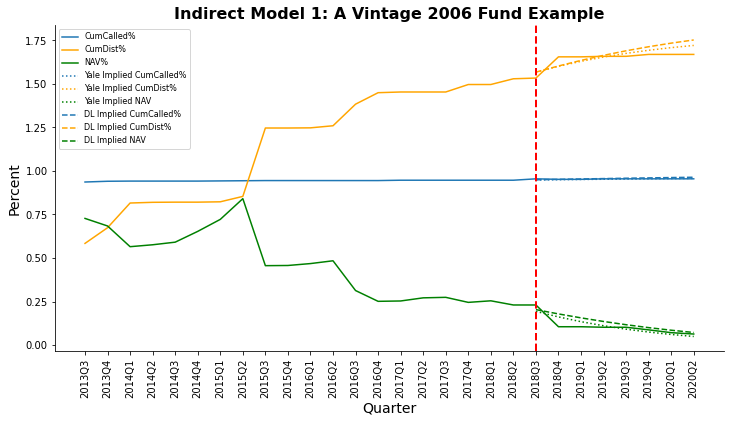}
   \end{minipage}\hfill
   \begin{minipage}{0.5\textwidth}
   \hspace*{0.5cm}
     \centering
     \includegraphics[scale = 0.35]{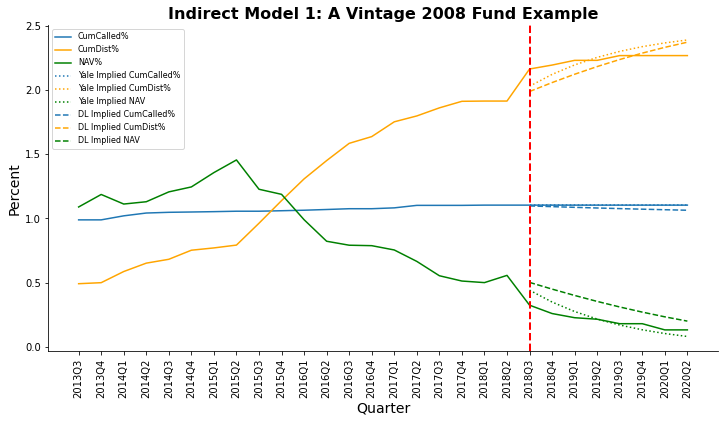}
   \end{minipage}
 \caption{Indirect Model 1 Predictions} \label{indirect_model1:pred}
\end{figure}

In our first model using indirect cash flow forecasting methodology, we estimate RC, G, and B parameters for two years ahead predictions using CC, DC, and RVC values introduced in Equations \ref{cc}, \ref{dc}, and \ref{rvc}. After tuning the hyperparameters, we use \texttt{GRU} model with three hidden layers that have 100, 75, and 100 hidden units, respectively. We use \textit{relu} activation function in order to introduce nonlinearity into the model. Using \textit{adam} optimizer, we train our model for 100 epochs. After the training, MSE value of the training set is 0.0018, and MSE value of the testing set is 0.0030. Figure \ref{indirect_model1:pred} illustrates two example prediction plots for the model, where we compare the model performance with actual ratios and benchmark model. According to the plots, the model performance highly overlaps with the benchmark model performance for 2006 vintage year fund. For 2008 vintage year fund example, there is a discrepancy between the model performance and benchmark model performance in terms of distribution and NAV predictions. Both examples show that the model is able to follow the trends of actual cash flows correctly, but the magnitude can be different for different fund and different rolling window performance. 

\FloatBarrier
In order to evaluate the impact of market shocks on the cash flows, we include five macroeconomic variables and two public market indexes into our model as input. Table \ref{table:macro} provides the list of the macroeconomic variables and market indices considered in this paper together with their data sources and frequencies. Accordingly, we use unemployment rate, GDP, yield, inflation, and gold price as macroeconomic variables. We also incorporate Russel 2000 and S\&P 500 indexes into the model.

\begin{table}[!htbp]
\vskip\baselineskip 
\begin{center}
\begin{tabular}{l|c|c} \hline
\textbf{Variables}& \textbf{Source} \tablefootnote{FRED: Federal Reserve Economic Data} & \textbf{Frequency} \\ \hline
    GDP &  FRED & Quarterly \\ 
    Unemployment Rate & FRED & Quarterly  \\ 
    CPI &FRED& Monthly \\ 
    Effective Yield & FRED & Daily \\ 
    Gold Price & FRED & Daily \\
    Russel 2000 & Yahoo Finance & Daily \\
    S\&P 500 & Yahoo Finance & Daily \\ \hline
\end{tabular}
\end{center}
\caption{Macroeconomic Variables and Market Indices }
\label{table:macro}
\end{table}

  We calculate percentage change of variables over a year, and then concatenate the percentages with cash flow data. We then pass the data into \texttt{GRU} with the same architecture. After training the model for 100 epochs, we obtain 0.0018 MSE value for training set and 0.0031 MSE value for testing set. Figure \ref{indirect} illustrates two prediction plots for 2006 and 2008 vintage year funds. The results show that model performance is almost the same with the previous model. Therefore, the impact of market environment is not visible when we introduce macroeconomic variables and public market indexes into the indirect cash flow forecasting model. 
\begin{figure}[!htbp]
\captionsetup{font=scriptsize,labelfont=scriptsize}
   \begin{minipage}{0.5\textwidth}
   \hspace*{-1.5cm}
     \centering
     \includegraphics[scale = 0.35]{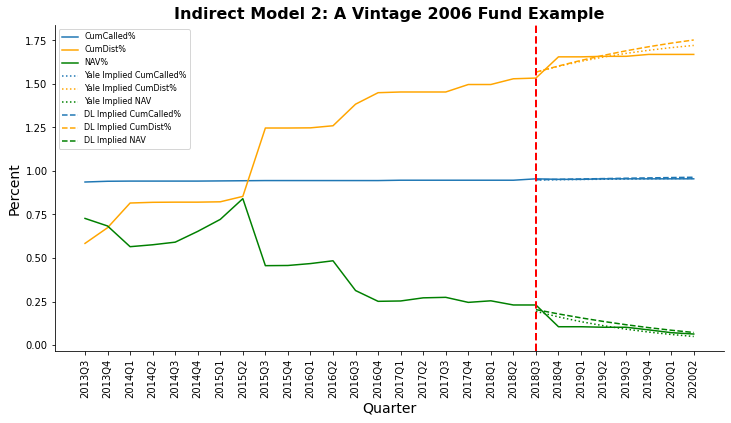}
   \end{minipage}\hfill
   \begin{minipage}{0.5\textwidth}
   \hspace*{0.5cm}
     \centering
     \includegraphics[scale = 0.35]{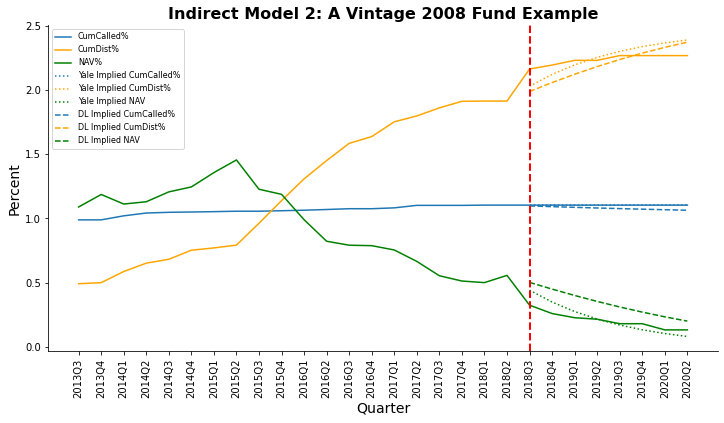}
   \end{minipage}
 \caption{Indirect Model 2 Predictions} \label{indirect_model2:pred}
\end{figure}
\FloatBarrier
\subsection{Direct Cash Flow Forecasting Results}
In this section, we experiment with the direct cash flow forecasting methodology, in which we do not rely on any assumptions. We employ both single vintage year models and multi-vintage year models with different input data to show how the model performs under different settings.

\subsubsection{Single vintage year models}

In this section, we utilize 2008 vintage year funds to build a preliminary prediction model and then evaluate how generic the model is. Our first model directly uses fund performance ratios Called \%, DPI \%, and RVPI \% to project future ratios.  

\begin{figure}[!htbp]
    \centering
    \includegraphics[scale = 0.4]{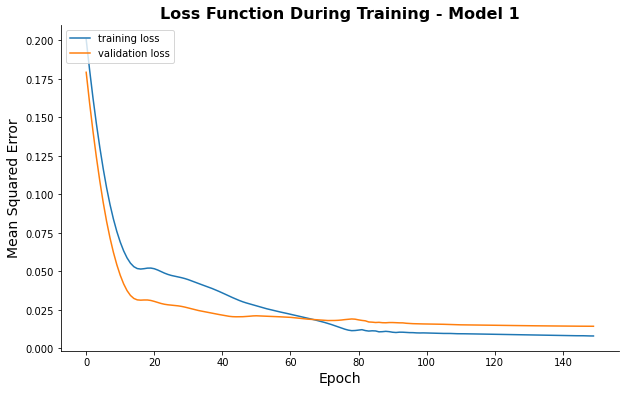}
    \caption{Training and validation loss evolution - model 1}
    \label{fig:model-1 loss}
\end{figure}
\FloatBarrier

In our first model, we don't use a rolling window approach. We consider 36 quarters lookback window, and then predict 13 quarters ahead. After tuning the hyperparameters, we use \texttt{GRU} model with three layers. First layer consists of 32 neurons with \textit{relu} function. We use RepeatVector to obtain prediction window dimension after the first layer. In the second layer, we again use 32 neurons with \textit{relu} activation function. The third layer consists of 16 neurons with \textit{sigmoid} activation function. The output is then wrapped in a time distributed dense layer to ensure the same weights are used for each time step and feature. Our model is trained for 150 epochs, and \textit{adam} optimizer is used to minimize mean squared error (MSE). According to the results, we obtained MSE value of 0.0084 for the training dataset and 0.0144 for the validation dataset. Figure \ref{fig:model-1 loss} illustrates the evolution of training and validation dataset mean squared errors. It shows that there is no overfitting problem while training. 

\begin{figure}[t]
\captionsetup{font=scriptsize,labelfont=scriptsize}
   \begin{minipage}{0.5\textwidth}
     \centering
     \includegraphics[scale = 0.32]{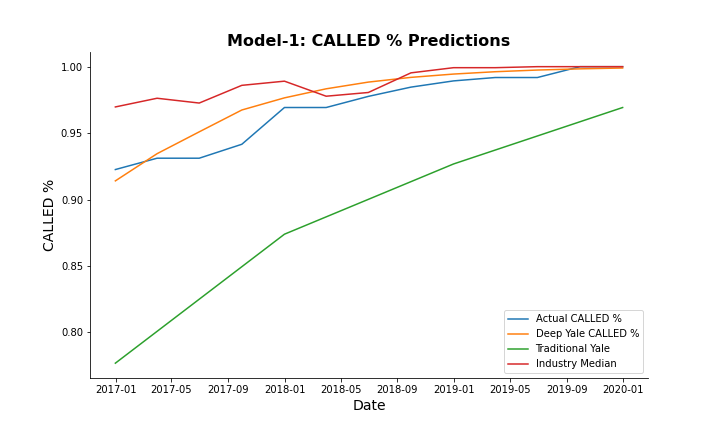}
   \end{minipage}\hfill
   \begin{minipage}{0.5\textwidth}
     \centering
     \includegraphics[scale = 0.32]{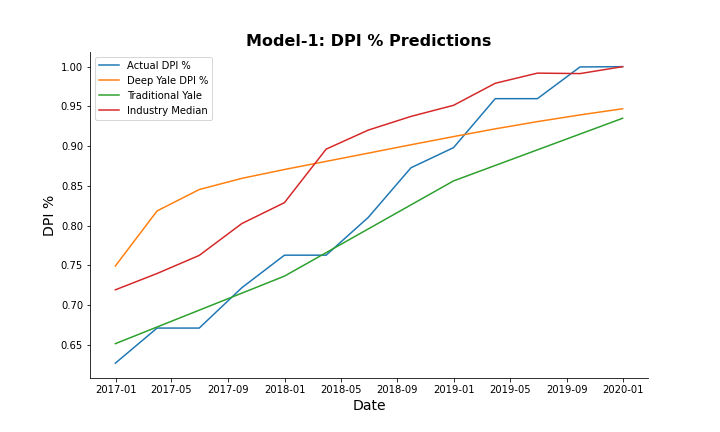}
   \end{minipage}\hfill
   \begin{center}
    \begin{minipage}{0.5\textwidth}
     \centering
     \includegraphics[scale = 0.32]{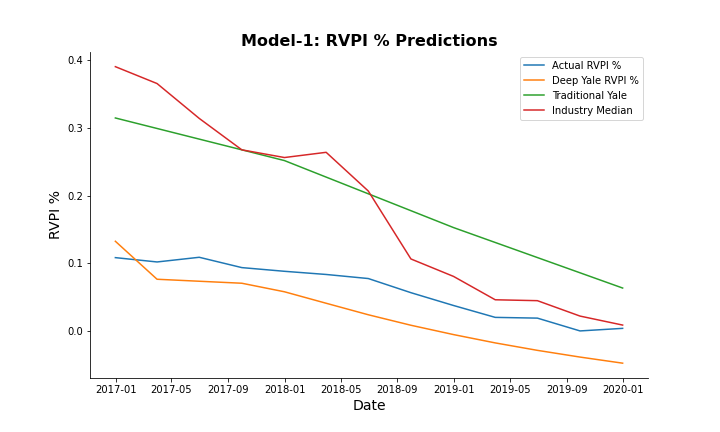}
   \end{minipage}
      \end{center}
 \caption{Model 1 Predictions} \label{model1:pred}
\end{figure}

Figure \ref{model1:pred} compares our predicted ratios with actual ratios and Yale model predicted ratios for a real fund. Accordingly, called \% predictions align with the actual values. Our predictions are better than Yale model predicted called \% values at capturing the pattern of called \%. Similarly, the pattern of RVPI \% predictions is very close to the pattern of actual RVPI \% values, and our predictions are much closer to the actual ratios compared to Yale model predicted RVPI \% ratios. However, Yale model predicted DPI \% values follow a closer pattern to the actual values compared to our model. 
\FloatBarrier

Our first model is able to follow the fund ratios, but our main goal is to predict exact values of contributions, distributions, and NAV values at each quarter. Therefore, we utilize CC, DC, and RVC values, which are shown in Equations \ref{cc}, \ref{dc}, and \ref{rvc}, in our second model to predict the contributions, distributions, and NAV values as percentages of commitment level. We can then multiply our predictions with the commitment level to obtain exact cash flows. 

\begin{figure}[!htbp]
    \centering
    \includegraphics[scale = 0.4]{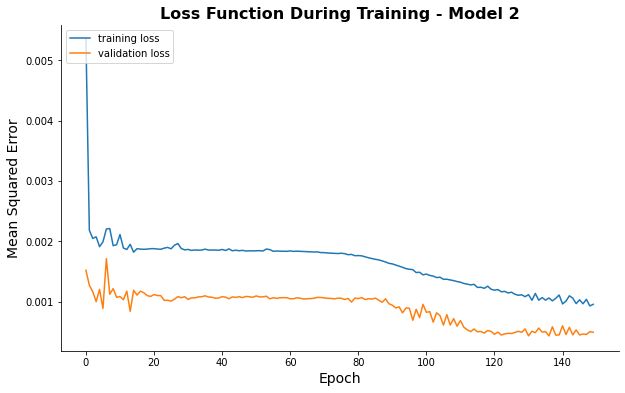}
    \caption{Training and validation loss evolution - model 2}
    \label{fig:model-2 loss}
\end{figure}

The second model uses 20 quarters lookback window and 8 quarters prediction window. After tuning the hyperparameters, we apply \texttt{GRU} model with three hidden layers. The first layer consists of 32 neurons with \textit{relu} activation function. We then use RepeatVector to create prediction window for each observation in training data. Second and third layers contain 32 neurons with \textit{relu} activation function and 16 neurons with \textit{sigmoid} activation function, respectively. The output is then wrapped in a time distributed dense layer together with exponential function to satisfy non-negativity of cash flows. The model is trained for 150 epochs, and \textit{adam} optimizer is used to minimize MSE. After training, MSE value of the training dataset is 0.00089, whereas MSE of validation dataset is 0.00049.  Figure \ref{fig:model-2 loss} shows the evolution of training and validation loss during the training. 

\FloatBarrier

\begin{figure}[!htbp]
\captionsetup{font=scriptsize,labelfont=scriptsize}
   \begin{minipage}{0.5\textwidth}
   \hspace*{-1.5cm}
     \centering
     \includegraphics[scale = 0.35]{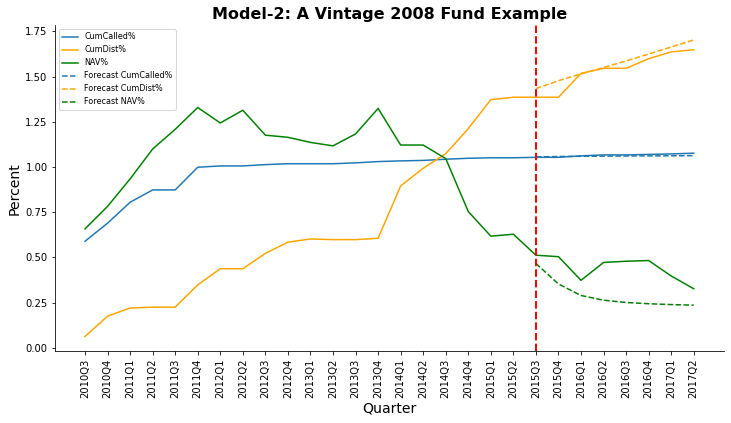}
   \end{minipage}\hfill
   \begin{minipage}{0.5\textwidth}
   \hspace*{0.5cm}
     \centering
     \includegraphics[scale = 0.35]{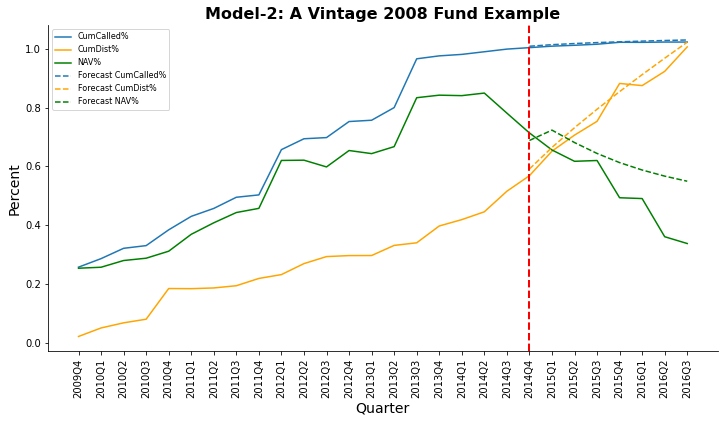}
   \end{minipage}
 \caption{Model 2 Predictions} \label{model2:pred}
\end{figure}

Figure \ref{model2:pred} provides two example prediction plots for the model that is built with 2008 vintage year fund data. Results show that predictions of quarter contributions and distributions are highly correlated with the realized quarter contribution and distribution values. NAV predictions are also aligned with the actual NAV values, but the relationship is weaker compared to the relationship between predictions and realizations of other cash flows. 

\FloatBarrier

\subsubsection{Multi-vintage year models}
Our initial models utilize only 2008 vintage year funds. However, the funds may perform different based on the time period they are active. Therefore, we integrate all vintage year funds together as explained earlier in \ref{Methodology-Yale}. For each fund, we first create rolling windows, and then pass lookback window data (20 quarters cash-flow data) to a chosen \texttt{RNN} model to project 8 quarters ahead for cash flows. 

\begin{figure}[!htbp]
\captionsetup{font=scriptsize,labelfont=scriptsize}
   \begin{minipage}{0.5\textwidth}
   \hspace*{-1.5cm}
     \centering
     \includegraphics[scale = 0.35]{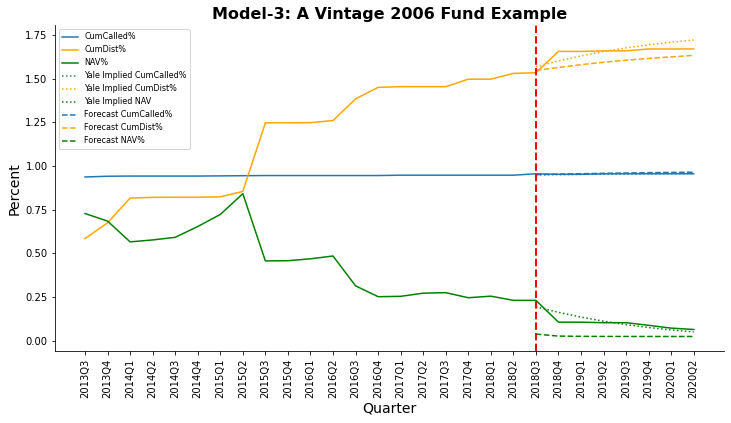}
   \end{minipage}\hfill
   \begin{minipage}{0.5\textwidth}
   \hspace*{0.5cm}
     \centering
     \includegraphics[scale = 0.35]{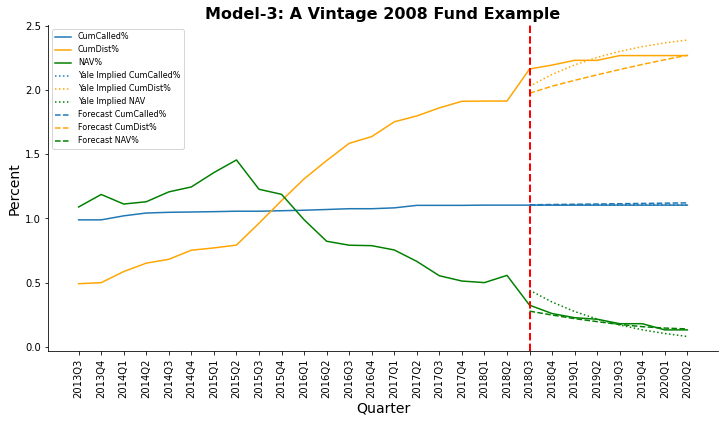}
   \end{minipage}
 \caption{Model 3 Predictions} \label{model3:pred}
\end{figure}

The first model in this section uses quarterly contributions, quarterly distributions, and NAV to build the prediction framework for cash flows. We implement \texttt{GRU} with three hidden layers after tuning hyperparameters. The model architecture is exactly the same with the Model 2 architecture from the previous section. We train our model for 250 epochs, and MSE values are reduced to 0.000203 for the training dataset and 0.0014 for the validation dataset. 
Figure \ref{model3:pred} provides one 2006 vintage fund example and  one 2008 vintage fund example from the validation dataset. According to both examples, contribution and distribution predictions are aligned with their corresponding realized values. Our model follows the trend of distributions better compared to the benchmark model. NAV predictions for 2008 vintage fund example is closely correlated with the actual values. On the other hand, benchmark model is better aligned with actual NAV values in 2006 vintage year example. In comparison to indirect cash flow forecasting models, the model is better at capturing cash flow patterns for both examples. 

\FloatBarrier

\begin{figure}[!htbp]
\captionsetup{font=scriptsize,labelfont=scriptsize}
   \begin{minipage}{0.5\textwidth}
   \hspace*{-1.5cm}
     \centering
     \includegraphics[scale = 0.35]{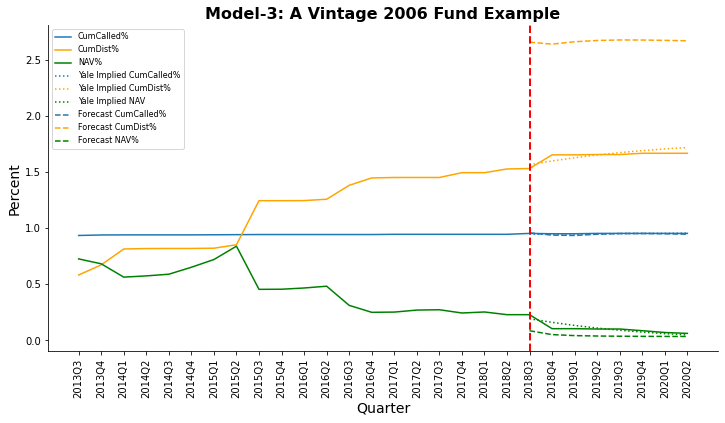}
   \end{minipage}\hfill
   \begin{minipage}{0.5\textwidth}
   \hspace*{0.5cm}
     \centering
     \includegraphics[scale = 0.35]{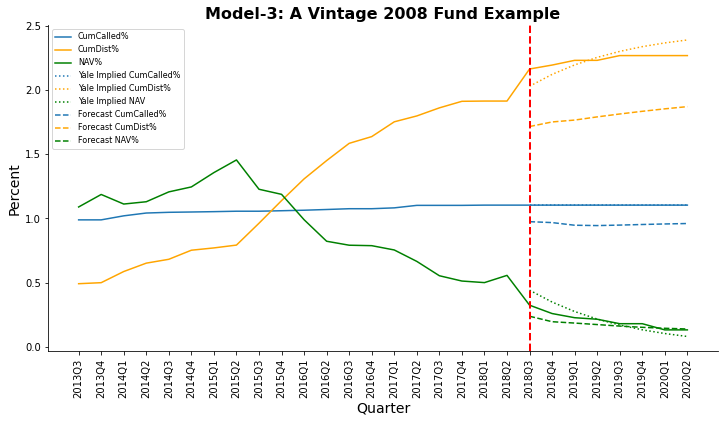}
   \end{minipage}
 \caption{Model 4 Predictions} \label{model4:pred}
\end{figure}

To compare the quality of predictions, we further build a model with cumulative contributions and cumulative distributions together with NAV values. We implement \texttt{GRU} with three hidden layers. The model architecture is same as previous two models except the last hidden layer contains 32 neurons in this model. Our model is trained for 100 epochs. The resulting loss function values are 0.0004 for the training dataset and 0.0028 for the validation dataset, which are higher compared to Model 3 loss function values. 
Figure \ref{model4:pred} shows the prediction plots for the same 2006 and 2008 vintage year funds. Predictions for contributions and NAV values are aligned with their corresponding realized values in 2006 vintage fund example. Similarly, NAV predictions are highly correlated with actual NAV values in 2008 vintage fund example. However, there is a discrepancy between predicted and realized cumulative distribution values in both examples. Therefore, we only consider quarterly cash flows for the following models. 

\FloatBarrier

To better capture the market environment influence on cash flows, we further introduce macroeconomic variables and public market indexes to our model. One year percentage changes of unemployment rate, GDP, yield, inflation, gold price, Russel 2000, and S\&P 500 values are concatenated with quarterly contributions, quarterly distribution, and NAV values, and then they are passed into \texttt{GRU} with three hidden layers. The model architecture is exactly the same with the architecture of Model 4. The model is then trained for 100 epochs, and the trained model gives 0.000093 and 0.0011 MSE values for training dataset and validation dataset, respectively. The loss function values indicate that use of macroeconomic variables and public market indexes in the model improve the performace of the prediction framework. 

\begin{figure}[!htbp]
\captionsetup{font=scriptsize,labelfont=scriptsize}
   \begin{minipage}{0.5\textwidth}
   \hspace*{-1.5cm}
     \centering
     \includegraphics[scale = 0.35]{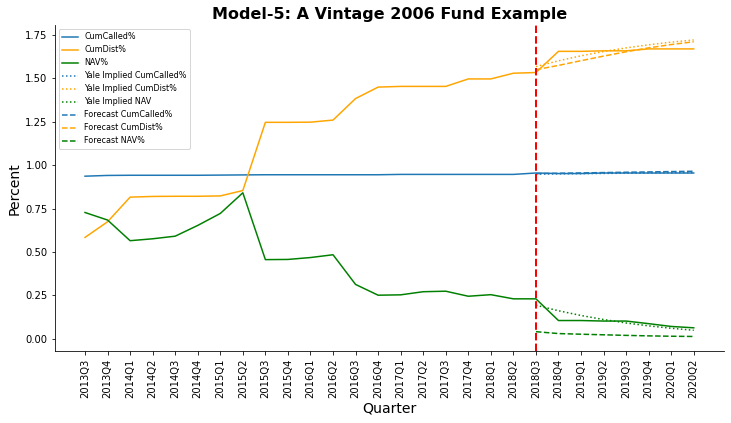}
   \end{minipage}\hfill
   \begin{minipage}{0.5\textwidth}
   \hspace*{0.5cm}
     \centering
     \includegraphics[scale = 0.35]{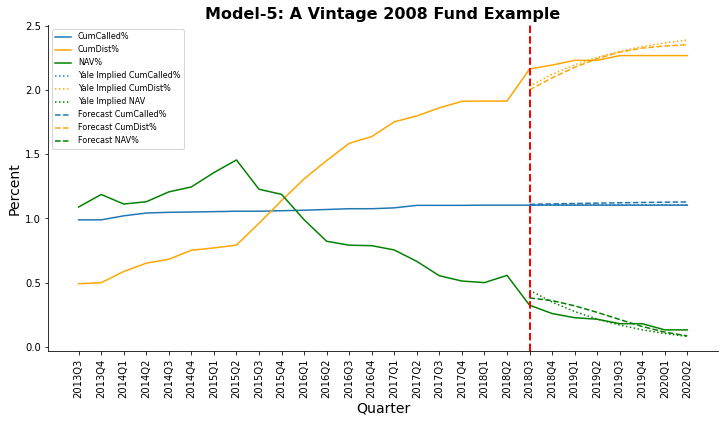}
   \end{minipage}
 \caption{Model 5 Predictions} \label{model5:pred}
\end{figure}

Figure \ref{model5:pred} illustrates example prediction plots for the same funds that are considered both in Model 3 and Model 4. Predicted contributions are aligned with realized contributions in both examples. 2006 vintage fund example shows a better fit for predicted distributions compared to 2008 vintage fund example. In contrast, predicted NAV values are more consistent with actual NAV values in 2008 vintage fund example. Compared to Figure \ref{model5:pred}, Figure \ref{model3:pred} still illustrates an overall better alignment with realized cash flows. 
\FloatBarrier

\section{Conclusion and Future Direction}\label{Conclusion-Yale}
All private equity funds follow a general J-curve pattern, yet there is an uncertainty in the timing and the size of the contributions and distributions. Although cash flow forecasting of private equity funds is a very important problem for investment purposes, the current literature is very limited in this field. In our study, we first convert our fund level data to equal length data by utilizing a sliding window approach, and then feed this data into an \texttt{LSTM/GRU} model to predict future cash flows. We introduce two different model frameworks, and compare their performance with quarterly adjusted Yale model. We further introduce a number of macroeconomic variables and market indices to our model to better capture the market environment. 

Our preliminary results indicate that the cash flows predicted using our models are aligned well with the actual cash flows. For indirect cash flow forecasting framework, our results indicate that macroeconomic variables and market indexes do not improve the model performance. For direct cash flow forecasting framework, both single vintage year and multi-vintage year models provide satisfactory results, yet multi-vintage year models are covering a broad range of funds from different vintage years. Although macroeconomic variables and market indexes improve the model performance for direct cash flow forecasting framework, we expect that a better selection of macroeconomic variables can increase the performance of cash flow forecasting further. 

Both model frameworks are mostly better at capturing future cash flow patterns compared to quarterly adjusted Yale model. When both model frameworks are compared to each other, indirect cash flow forecasting model still relies on the certain assumptions such as life of the fund, and the model parameter estimations. It provides better alignment compared to benchmark model because the parameters are dynamically adjusted for the future cash flows, but direct cash flow forecasting framework results show that Model 3, 4, and 5 are more flexible at following cash flow patterns. Implementing a parameter estimation model before deep neural networks adds more noise into the data, and the model performance is impacted. Furthermore, direct cash flow forecasting framework is more suitable to measure the market shocks because it reacts to the market change.

As a future work, we plan to improve our model architecture by benefiting from attention-based recurrent neural networks and generative adversarial networks (\texttt{GAN}s). We aim at extending our prediction window without sacrificing the model's ability to capture the patterns of cash flows. A more thorough comparative and backtesting results and a stress-testing framework will be added later as a future extension.

\clearpage

\printbibliography


\end{document}